\documentclass[prb,showpacs,twocolumn]{revtex4-1}
\usepackage{graphicx}
\begin{document}

\title{Magnetotransport properties in a noncentrosymmetric itinerant magnet Cr$_{11}$Ge$_{19}$}

\author{N. Jiang$^{1}$}
\email{jiangnan@mail.ecc.u-tokyo.ac.jp}
\author{Y. Nii$^{1}$}
\author{R. Ishii$^{2}$}
\author{Z. Hiroi$^{2}$}
\author{Y. Onose$^{1}$}
\affiliation{$^{1}$Department of Basic Science, University of Tokyo 153-8902, Japan\\
$^{2}$Institute for Solid State Physics, University of Tokyo, Kashiwa, Chiba 277-8581, Japan}

\date{\today}

\begin{abstract}
 We have investigated anomalous Hall effect and magnetoresistance in a noncentrosymmetric itinerant magnet Cr$_{11}$Ge$_{19}$. While the temperature- and magnetic-field-dependent anomalous Hall conductivity is just proportional to the magnetization above 30 K, it is more enhanced in the lower temperature region. The magnitude of negative magnetoresistance begins to increase toward low temperature around 30 K. The anisotropic magnetoresistance emerges at similar temperature. Because there is no anomaly in the temperature dependence of magnetization around 30 K, the origin of these observations in transport properties is ascribed to some electronic structure with the energy scale of 30 K. We speculate this is caused by the spin splitting due to breaking of spatial inversion symmetry.
\end{abstract}


\maketitle

\section{INTRODUCTION}
 Recently, noncentrosymmetric magnets have been studied extensively, because noncollinear spin textures are emergent in real and momentum spaces. In this class of materials, there is the  spatially uniform component of the Dzyaloshinsky-Moriya interaction, which produces, in some cases, real space spin textures as a result of competition with the exchange interaction. The skyrmion lattice\cite{skyrmion1,skyrmion2} and the chiral soliton lattice\cite{CRL} are examples of them. In particular, in skyrmion lattice, electrons acquire Berry's phase due to the topological nature of spin structure, which gives rise to nontrivial Hall signal denoted as topological Hall effect\cite{THE}. 
In noncentrosymmetric metals and semiconductors, the electronic bands are spin splitted along the principal axis dependent on the momentum due to the spin orbit interaction. The examples are the Rashba and Dresselhaus effects\cite{Rashba,Dresselhaus}. In the itinerant magnets without spatial inversion symmetry, the coexistence of spin orbit driven spin splitting and localized magnetic moments gives raise to novel magnetoelectric phenomena. For example, the spin dependent scattering caused by magnetic moments in a spin splitted band gives raise to large anisotropic magnetoresistance (AMR)\cite{GaAsAMR}. The anomalous Hall effect induced by momentum space Berry phase is also anticipated in a exchange biased Rashba band\cite{AHE1,AHE2}. The magnetic moment can be controlled by the spin orbit torque produced by the electric current in a spin splitted band\cite{SOT1}. Because these novel phenomena depend on the crystal and magnetic structures, further studies of various noncentrosymmetric magnets seem quite important for the deep understanding of related physics. 

Here we study a noncentrosymmetric itinerant magnet Cr$_{11}$Ge$_{19}$ with a space group $P\overline{4}n2$\cite{CrGe1,CrGe2,CrGe3,CrGe4}. It shows a magnetic transition below about 90 K. The magnetization curve below Curie temperature $T_{\rm C}$ is consistent with the picture of anisotropic ferromagnet with the easy axis along $c$-axis. The effect of breaking of inversion symmetry has not been reported so far. 
 In this paper, we have investigated electric and magnetic properties in a single crystal of Cr$_{11}$Ge$_{19}$ in order to exploit the effect of breaking of inversion symmetry. We observe enhancement of anomalous Hall conductivity and the magnitude and anisotropy of magnetoresistance around 30 K. We speculate these anomalies are caused by the spin splitting due to inversion symmetric breaking.

\section{EXPERIMENTAL DETAILS}
 Single crystals of Cr$_{11}$Ge$_{19}$ were prepared by the self flux method. Mixed powder of Cr and Ge with the ratio of 1:4 were loaded in a alumina crucible. The crucible  was sealed together with zirconium getter in a silica tube under vacuum. The tube was placed in a Bridgman furnace and heated to 1100 $^\circ$C in 20 hours and kept for 60 hours. After cooled to 1000 $^\circ$C in 5 hours, the tube was moved toward low temperature region in temperature gradient ($\sim$ 1-5 $^\circ$C/mm) at the rate of 0.5 mm per hour for 300 hours. The crystals with the dimensions of several millimeter were obtained. The single crystallinity was confirmed by Laue X-ray diffraction. Magnetization was measured in a quantum design magnetic property measurement system (MPMS). The longitudinal and Hall resistivities were measured by conventional four probe method in a superconducting magnet (Oxford instruments Spectromag).

\section{RESULTS AND DISCUSSION}

\begin{figure}
\includegraphics[width=5cm]{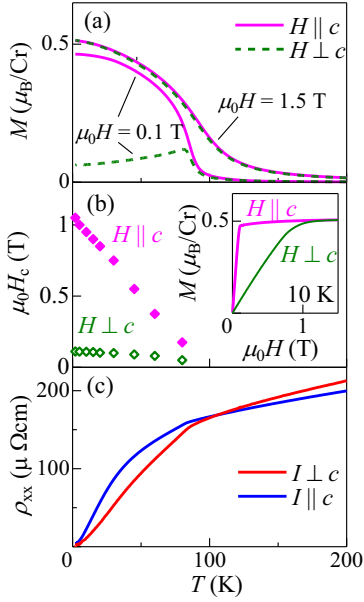}
\caption{Temperature dependence of  (a) magnetization at 0.1 T and 1.5 T for $H \bot c$ and $H || c$, (b) saturation magnetic fields, and $H || c$ and (c) zero field resistivity for $I \bot c$ and $I || c$. Inset in (b) shows magnetization curves for $H || c$ and $H \bot c$ at 10 K.} 
\end{figure}

 Figure 1 (a) shows the temperature dependence of the magnetization $M$ in magnetic fields $H$ of 0.1 T and 1.5 T for $H\bot c$ and $H||c$. The magnetization begins to rapidly increase around 85 K, suggesting the ferromagnetic transition. The magnetization along $c$-axis is larger than that along $c$-plane at 0.1 T, being consistent with the previous study\cite{CrGe1}. At 1.5 T, the anisotropy of magnetization becomes very small. In the inset of Fig. 1 (b), we show the magnetization curves for $H \bot c$ and $H || c$ at 10 K. The magnetization curves for $H \bot c$ and $H || c$ saturate around 1 T and 0.1 T, respectively. After saturation, the magnitudes for $H \bot c$ and $H || c$ are almost identical. Figure 1 (b) shows the temperature dependence of the saturation magnetic fields $H_{\rm c}$. $H_{\rm c}$ for $H \bot c$ is larger than that for $H || c$, and both increase almost linearly below $T_{\rm C}$. In Fig. 1 (c), we show the temperature dependence of resistivities $\rho$ measured with electric currents $I$ perpendicular and parallel to $c$-axis ($I \bot c$ and $I || c$). For both the configurations, the resistivity shows a clear kink structure at $T_{\rm C}$ while the temperature dependence is small in the higher temperature region. Below $T_{\rm C}$, the resistivity decrease rapidly due to the suppression of magnetic fluctuation, and the anisotropy is enhanced. The residual resistivities for $I \bot c$ and $I || c$ were 1.5 and 3.3 $\mu  \Omega$cm, respectively, indicating the high quality of present crystals.

\begin{figure}
\includegraphics[width=7.5cm]{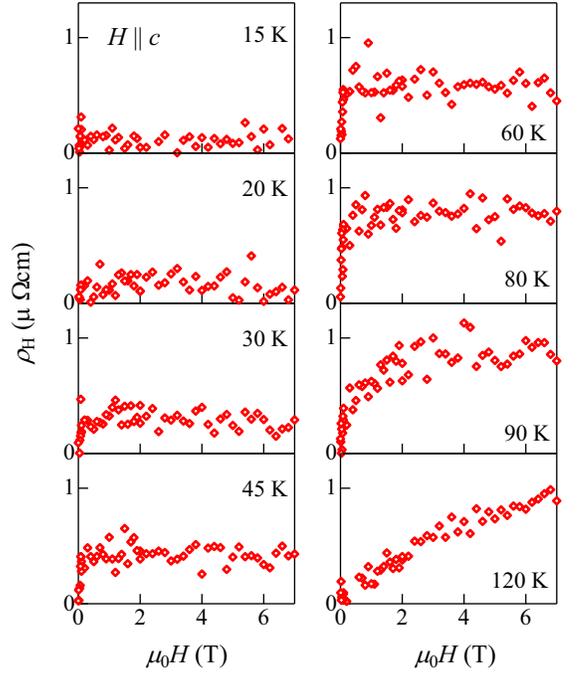}
\caption{Magnetic field dependence of Hall resistivity at various temperatures.} 
\end{figure}

\begin{figure}
\includegraphics[width=5cm]{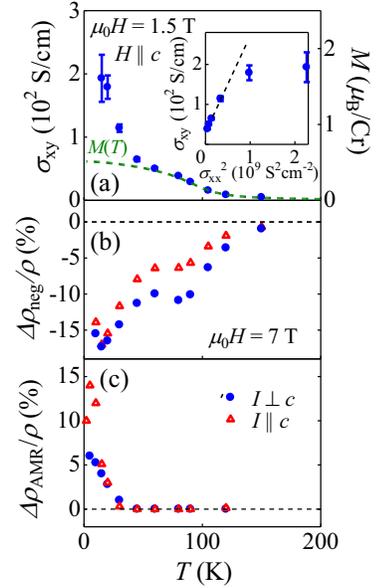}
\caption{(a) Temperature dependence of Hall conductivity (red dots) and magnetization (green dashed line) around 1.5 T. The inset shows $\sigma_{\rm AH}$ potted against $\sigma_{xx}^{2}$. The black dashed line shows the $\sigma_{xx}$ dependence expected in the case of coexistence of skew scattering and intrinsic mechanisms (see main text). (b) Temperature dependences of $\Delta\rho_{\rm neg}/\rho$ for $I \bot c$ and $I || c$. (c) Temperature dependences of $\Delta\rho_{\rm AMR}/\rho$ for $I \bot c$ and $I || c$.} 
\end{figure}

 In Fig. 2, we show the magnetic field dependence of Hall resistivity $\rho_{\rm H}$ at respective temperatures. While the Hall resistivity shows linear magnetic field dependence at 120 K, it tends to saturate in the high field and the lower temperature region. Below  $T_{\rm C}$ the spontaneous component of Hall resistivity emerges. It decreases with decreasing temperature from $T_{\rm C}$. The magnetic field dependence of Hall resistivity is quite similar to that of magnetization. In the low temperature and high field region, in which the magnetization is saturated, the Hall resistivity does not show large magnetic field dependence. These suggest the dominant contribution of anomalous component.
In Fig. 3 (a), we plot the temperature dependence of the Hall conductivity $\sigma_{xy} = \rho_{\rm H} / (\rho_{xx}^{2} + \rho_{\rm H}^{2})$  around 1.5 T for $H || c$. To be strict, we average the data between 0.3 T to 2.6 T in order to precisely estimate $\sigma_{xy}$. The error bar represents the standard error. We also plot the temperature dependence of the magnetization at 1.5 T for comparison. Above 30 K, the temperature dependence of $\sigma_{xy}$ is parallel with that of magnetization. This is consistent with intrinsic anomalous Hall effect\cite{Nagaosa2010}. Nevertheless, $\sigma_{xy}$ becomes larger than that expected from the scaling behavior below 30 K. Such an enhancement of $\sigma_{xy}$ might be caused by the emergence of skew scattering contribution. If this is the case, $\sigma_{xy}$ should satisfy the relation of $\sigma_{xy} = \alpha\sigma_{xx}^{2}+ \beta$, where $\sigma_{xx} = 1 / \rho_{xx}$, and $\alpha$ and $\beta$ are constants\cite{properscalingAHE}. However, in the present case, $\sigma_{xy}$ does not follow this relation as shown in the inset of Fig. 3 (a) inset. As discussed below, the magnetoresistance also shows anomalies in similar temperature range.

\begin{figure}
\includegraphics[width=5cm]{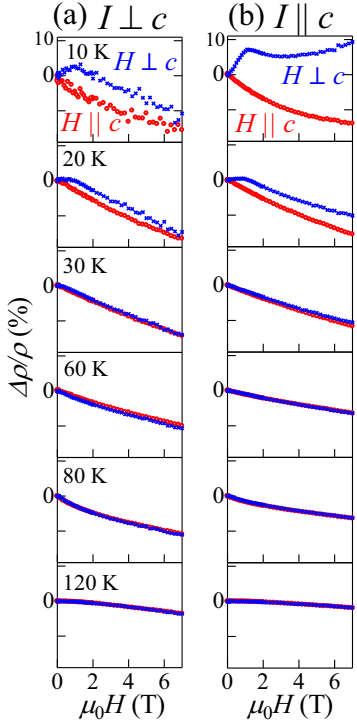}
\caption{(a), (b) Magnetic field dependence of magnetoresistance $\Delta\rho/\rho = (\rho_{xx}(H)- \rho_{xx}(H=0))/ \rho_{xx}(H=0)$ for (a) $I \bot c$ and (b) $I || c$ in the magnetic fields parallel and perpendicular to $c$-axis at various temperatures.} 
\end{figure}

\begin{figure}
\includegraphics[width=8.5cm]{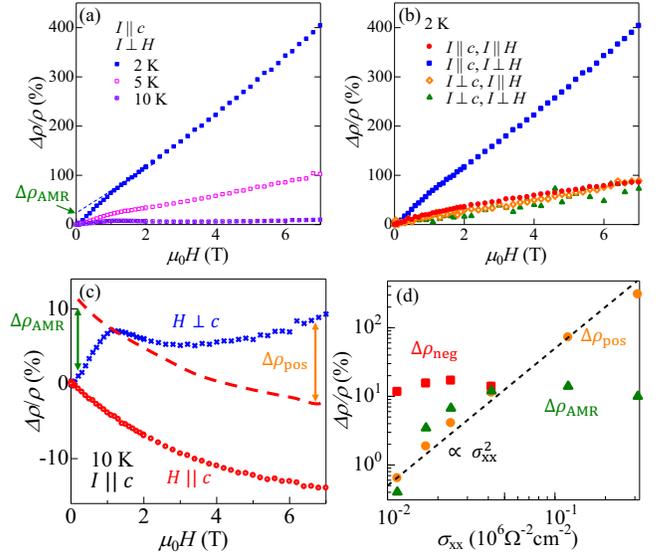}
\caption{(a) Magnetic field dependence of $\Delta\rho/\rho$ for $I || c$ and $H \bot c$ at 2 K, 5 K, and 10 K. (b) Magnetic field dependence of $\Delta\rho/\rho$ at 2K with various configurations of current and magnetic field. (c) Magnetic field dependence of $\Delta\rho/\rho$ for $I || c$ in the magnetic fields parallel and perpendicular to $c$-axis at 10 K. Dashed line shows vertically shifted $\Delta\rho/\rho$ for $H || c$. $\Delta\rho_{\rm pos}$ and $\Delta\rho_{\rm AMR}$ indicate the estimated components of positive magnetoresistance and AMR. (d) $\Delta\rho_{\rm pos}$ at 7 T, $\Delta\rho_{\rm neg}$ at 7 T, and $\Delta\rho_{\rm AMR}$ for $I || c$ plotted against $\sigma_{xx}$.} 
\end{figure}

 We show the magnetic field dependence of magnetoresistance  $\Delta\rho/\rho = (\rho_{xx}(H)- \rho_{xx}(H=0))/ \rho_{xx}(H=0)$ in magnetic fields parallel and perpendicular to $c$-axis in Fig. 4 (a) for $I \bot c$ and in Fig. 4 (b) for $I || c$. The magnetoresistance data was measured in the transverse configuration ($I \bot H$) except for the $H || c$ and $I || c$ configuration. In conventional itinerant ferromagnets, negative magnetoresistance is observed around $T_{\rm C}$ due to spin scattering\cite{NMR}. It usually decreases with decreasing temperature. Such a negative magnetoresistance is certainly observed around $T_{\rm C}$ also in the present case. In this temperature range, the magnetoresistance is isotropic with respect to the magnetic field direction. With decreasing temperature from $T_{\rm C}$, the magnetoresistance decrease at first. However, below 30 K the magnetoresistance begins to increase again. The magnetoresistances for $H \bot c$ and $H || c$ become different at lower temperature for both the $I || c$ and $I \bot c$ configurations. The anisotropy is increased in the low field region and saturated around 1 T. These observations suggest that the resistivity depends on the direction of the magnetization, which is ascribed to anisotropic magnetoresistance induced by spin orbit interaction\cite{AMR3d}.
Incidentally, the additional $H$-dependent positive magnetoresistance is observed in the high field region at 10 K for $I || c$ and $H \bot c$. The positive magnetoresistance steeply increases with decreasing temperature from 10 K as shown in Fig. 5 (a). In  Fig. 5 (b) we show the magnetoresistance for various current and field configurations at 2 K. While the positive magnetoresistance was observed at 2 K in all the configuration, that for $I || c$ and $H \bot c$ is most enhanced.

Phenomenologically, the observed magnetoresistance is composed of three components as follows; 
\begin{equation}
\Delta\rho/\rho = \Delta\rho_{\rm pos}  + \Delta\rho_{\rm neg}  + \Delta\rho_{\rm AMR}.
\end{equation}
$\Delta\rho_{\rm pos}$ is positive magnetoresistance enhanced in the low temperature region mainly in the configuration of $I || c$ and $I \bot H$. $\Delta\rho_{\rm neg}$ is negative magnetoresistance. This component seems isotropic. $\Delta\rho_{\rm AMR}$ is anisotropic magnetoresistance dependent on the direction of magnetization. We estimate these components for the $I || c$ magnetoresistance above 10 K as illustrated in Figs. 5 (c). We assume the isotropy of $\Delta\rho_{\rm neg}$ and field independence above $H_{\rm c}$ for $\Delta\rho_{\rm AMR}$. We also assume that $H || c$ data above 10 K is dominated by the $\Delta\rho_{\rm neg}$ component. Therefore $\Delta\rho_{\rm AMR}$ can be estimated by the extrapolation of $H \bot c$ data above $H_{\rm c}$ to zero field. The $\Delta\rho_{\rm pos}$ component is estimated as $\Delta\rho/\rho$ for $H \bot c$ subtracted by $\Delta\rho_{\rm AMR}$ and $\Delta\rho_{\rm neg}$. Below 5 K, we assume $\Delta\rho_{\rm neg}$ is negligible compared with the other components and $\Delta\rho_{\rm AMR}$ is estimated by the linear extrapolation (see Fig. 5 (a)). The estimated $\Delta\rho_{\rm neg}$ at 7 T, $\Delta\rho_{\rm pos}$ at 7 T, and $\Delta\rho_{\rm AMR}$ are plotted against $\sigma_{xx}$ in Fig. 5 (d). $\Delta\rho_{\rm pos}$ steeply increased roughly as $\sigma_{xx}^{2}$, while the increase of $\Delta\rho_{\rm AMR}$ and $\Delta\rho_{\rm neg}$ are not so significant. Such a positive magnetoresistance enhanced in the low resistive regime is usually ascribed to the orbital magnetoresistance. The orbital magnetoresistance is caused by Lorentz force and becomes significant when $\omega_{\rm c}\tau$ is large ($\omega_{\rm c}$ and $\tau$ are cyclotron frequency and relaxation time, respectively). The orbital magnetoresistance satisfies Kohler's rule, in which magnetoresistance is scaled by $H / \rho_{xx}(H = 0)$. While the scaling behavior cannot be directly confirmed in this case because of coexistence of other type magnetoresistance, the power law $\sigma_{xx}$ dependence is consistent with the Kohler's rule. On the other hand, the low temperature increase of $\Delta\rho_{\rm AMR}$ and $\Delta\rho_{\rm neg}$ can not be ascribed to the orbital magnetoresistance, because the $\sigma_{xx}$ dependences are quite gradual compared with $\Delta\rho_{\rm pos}$, and because the magnetic field dependences are different from that expected for orbital magnetoresistance ($\Delta\rho_{\rm AMR} \propto H^{0}$ in high field and $\Delta\rho_{\rm neg} <$ 0).

Apart from orbital magnetoresistance, let us discuss the low temperature anomalies of magnetotransport properties. 
We show the temperature dependence of $\Delta\rho_{\rm neg}$ and $\Delta\rho_{\rm AMR}$ in Fig. 3 (b) and (c). The $I \bot c$ data shown in this figure is estimated similarly to the case of $I || c$. $\Delta\rho_{\rm AMR}$ begins to increase with decreasing temperature below 30 K. Similar low temperature enhancements of AMR were reported for some itinerant ferromagnets, which is caused by the crossover of dominant scattering source from electron phonon type to impurity type\cite{AMRFe,AMRSmit}. Because the impurity scattering induces the large AMR compared with electron phonon scattering, the magnitude of AMR is increased as the resistivity approaches the $T$ = 0 value. However, in our case the resistivity around the onset of AMR is much larger than the residual resistivity. Therefore, there should be another origin of temperature dependence. Around the onset temperature,  $\Delta\rho_{\rm neg}$ begins to increase and the anomalous Hall conductivity is enhanced compared with magnetization. These three anomalies seem to have a common origin. The magnetization and the saturation magnetic fields does not show any anomaly around this temperature as shown in Figs. 1 (a) and (b). Therefore, the above mentioned transport anomalies should originate from some electronic structure with the energy scale of 30 K. Because Cr$_{11}$Ge$_{19}$ is a noncentrosymmetric magnet, the spin splitting caused by spin orbit interaction seems the most plausible origin. As mentioned in introduction, in ferromagnets without inversion symmetry, anomalous Hall effect can be induced by the Berry curvature of spin splitted band. The anisotropic magnetoresistance can also be induced by spin dependent scattering by localized moment. The origin of negative magnetoresistance may be caused by the magnetic field induced change of spin splitted bands. However, at present, this scenario cannot be examined extensively because of the lack of the information regarding spin-splitted band structure. In order to confirm this scenario relativistic band calculation is needed.

\section{SUMMARY}
 We have uncovered several magnetotransport anomalies below around 30 K. The negative magnetoresistance and anomalous Hall conductivity are largely enhanced in this temperature region. Anisotropic magnetoreisistance is also observed in a similar temperature region. We speculate these observations are caused by the spin splitting due to the breaking of spatial inversion symmetry. The investigation of relativistic energy band is indispensable for understanding of spin dependent transport properties in Cr$_{11}$Ge$_{19}$.

\begin{acknowledgments}
 This work was supported in part by Grants-in-Aid for scientific research (Grants no.17H05176, no.16H04008, and no.15K21622) for the Japan Society for the promotion of Science.
\end{acknowledgments}

\bibliography{Cr11Ge19ref}

\begin{thebibliography}{20}%
\makeatletter
\providecommand \@ifxundefined [1]{%
 \@ifx{#1\undefined}
}%
\providecommand \@ifnum [1]{%
 \ifnum #1\expandafter \@firstoftwo
 \else \expandafter \@secondoftwo
 \fi
}%
\providecommand \@ifx [1]{%
 \ifx #1\expandafter \@firstoftwo
 \else \expandafter \@secondoftwo
 \fi
}%
\providecommand \natexlab [1]{#1}%
\providecommand \enquote  [1]{``#1''}%
\providecommand \bibnamefont  [1]{#1}%
\providecommand \bibfnamefont [1]{#1}%
\providecommand \citenamefont [1]{#1}%
\providecommand \href@noop [0]{\@secondoftwo}%
\providecommand \href [0]{\begingroup \@sanitize@url \@href}%
\providecommand \@href[1]{\@@startlink{#1}\@@href}%
\providecommand \@@href[1]{\endgroup#1\@@endlink}%
\providecommand \@sanitize@url [0]{\catcode `\\12\catcode `\$12\catcode
  `\&12\catcode `\#12\catcode `\^12\catcode `\_12\catcode `\%12\relax}%
\providecommand \@@startlink[1]{}%
\providecommand \@@endlink[0]{}%
\providecommand \url  [0]{\begingroup\@sanitize@url \@url }%
\providecommand \@url [1]{\endgroup\@href {#1}{\urlprefix }}%
\providecommand \urlprefix  [0]{URL }%
\providecommand \Eprint [0]{\href }%
\providecommand \doibase [0]{http://dx.doi.org/}%
\providecommand \selectlanguage [0]{\@gobble}%
\providecommand \bibinfo  [0]{\@secondoftwo}%
\providecommand \bibfield  [0]{\@secondoftwo}%
\providecommand \translation [1]{[#1]}%
\providecommand \BibitemOpen [0]{}%
\providecommand \bibitemStop [0]{}%
\providecommand \bibitemNoStop [0]{.\EOS\space}%
\providecommand \EOS [0]{\spacefactor3000\relax}%
\providecommand \BibitemShut  [1]{\csname bibitem#1\endcsname}%
\let\auto@bib@innerbib\@empty
\bibitem [{\citenamefont {M{\"u}hlbauer}\ \emph {et~al.}(2009)\citenamefont
  {M{\"u}hlbauer}, \citenamefont {Binz}, \citenamefont {Jonietz}, \citenamefont
  {Pfleiderer}, \citenamefont {Rosch}, \citenamefont {Neubauer}, \citenamefont
  {Georgii},\ and\ \citenamefont {B{\"o}ni}}]{skyrmion1}%
  \BibitemOpen
  \bibfield  {author} {\bibinfo {author} {\bibfnamefont {S.}~\bibnamefont
  {M{\"u}hlbauer}}, \bibinfo {author} {\bibfnamefont {B.}~\bibnamefont {Binz}},
  \bibinfo {author} {\bibfnamefont {F.}~\bibnamefont {Jonietz}}, \bibinfo
  {author} {\bibfnamefont {C.}~\bibnamefont {Pfleiderer}}, \bibinfo {author}
  {\bibfnamefont {A.}~\bibnamefont {Rosch}}, \bibinfo {author} {\bibfnamefont
  {A.}~\bibnamefont {Neubauer}}, \bibinfo {author} {\bibfnamefont
  {R.}~\bibnamefont {Georgii}}, \ and\ \bibinfo {author} {\bibfnamefont
  {P.}~\bibnamefont {B{\"o}ni}},\ }\href {\doibase 10.1126/science.1166767}
  {\bibfield  {journal} {\bibinfo  {journal} {Science}\ }\textbf {\bibinfo
  {volume} {323}},\ \bibinfo {pages} {915} (\bibinfo {year}
  {2009})}\BibitemShut {NoStop}%
\bibitem [{\citenamefont {Yu}\ \emph {et~al.}(2010)\citenamefont {Yu},
  \citenamefont {Onose}, \citenamefont {Kanazawa}, \citenamefont {Park},
  \citenamefont {Han}, \citenamefont {Matsui}, \citenamefont {Nagaosa},\ and\
  \citenamefont {Tokura}}]{skyrmion2}%
  \BibitemOpen
  \bibfield  {author} {\bibinfo {author} {\bibfnamefont {X.}~\bibnamefont
  {Yu}}, \bibinfo {author} {\bibfnamefont {Y.}~\bibnamefont {Onose}}, \bibinfo
  {author} {\bibfnamefont {N.}~\bibnamefont {Kanazawa}}, \bibinfo {author}
  {\bibfnamefont {J.}~\bibnamefont {Park}}, \bibinfo {author} {\bibfnamefont
  {J.}~\bibnamefont {Han}}, \bibinfo {author} {\bibfnamefont {Y.}~\bibnamefont
  {Matsui}}, \bibinfo {author} {\bibfnamefont {N.}~\bibnamefont {Nagaosa}}, \
  and\ \bibinfo {author} {\bibfnamefont {Y.}~\bibnamefont {Tokura}},\
  }\href@noop {} {\bibfield  {journal} {\bibinfo  {journal} {Nature}\ }\textbf
  {\bibinfo {volume} {465}},\ \bibinfo {pages} {901} (\bibinfo {year}
  {2010})}\BibitemShut {NoStop}%
\bibitem [{\citenamefont {Togawa}\ \emph {et~al.}(2012)\citenamefont {Togawa},
  \citenamefont {Koyama}, \citenamefont {Takayanagi}, \citenamefont {Mori},
  \citenamefont {Kousaka}, \citenamefont {Akimitsu}, \citenamefont {Nishihara},
  \citenamefont {Inoue}, \citenamefont {Ovchinnikov},\ and\ \citenamefont
  {Kishine}}]{CRL}%
  \BibitemOpen
  \bibfield  {author} {\bibinfo {author} {\bibfnamefont {Y.}~\bibnamefont
  {Togawa}}, \bibinfo {author} {\bibfnamefont {T.}~\bibnamefont {Koyama}},
  \bibinfo {author} {\bibfnamefont {K.}~\bibnamefont {Takayanagi}}, \bibinfo
  {author} {\bibfnamefont {S.}~\bibnamefont {Mori}}, \bibinfo {author}
  {\bibfnamefont {Y.}~\bibnamefont {Kousaka}}, \bibinfo {author} {\bibfnamefont
  {J.}~\bibnamefont {Akimitsu}}, \bibinfo {author} {\bibfnamefont
  {S.}~\bibnamefont {Nishihara}}, \bibinfo {author} {\bibfnamefont
  {K.}~\bibnamefont {Inoue}}, \bibinfo {author} {\bibfnamefont {A.~S.}\
  \bibnamefont {Ovchinnikov}}, \ and\ \bibinfo {author} {\bibfnamefont
  {J.}~\bibnamefont {Kishine}},\ }\href {\doibase
  10.1103/PhysRevLett.108.107202} {\bibfield  {journal} {\bibinfo  {journal}
  {Phys. Rev. Lett.}\ }\textbf {\bibinfo {volume} {108}},\ \bibinfo {pages}
  {107202} (\bibinfo {year} {2012})}\BibitemShut {NoStop}%
\bibitem [{\citenamefont {Neubauer}\ \emph {et~al.}(2009)\citenamefont
  {Neubauer}, \citenamefont {Pfleiderer}, \citenamefont {Binz}, \citenamefont
  {Rosch}, \citenamefont {Ritz}, \citenamefont {Niklowitz},\ and\ \citenamefont
  {B\"oni}}]{THE}%
  \BibitemOpen
  \bibfield  {author} {\bibinfo {author} {\bibfnamefont {A.}~\bibnamefont
  {Neubauer}}, \bibinfo {author} {\bibfnamefont {C.}~\bibnamefont
  {Pfleiderer}}, \bibinfo {author} {\bibfnamefont {B.}~\bibnamefont {Binz}},
  \bibinfo {author} {\bibfnamefont {A.}~\bibnamefont {Rosch}}, \bibinfo
  {author} {\bibfnamefont {R.}~\bibnamefont {Ritz}}, \bibinfo {author}
  {\bibfnamefont {P.~G.}\ \bibnamefont {Niklowitz}}, \ and\ \bibinfo {author}
  {\bibfnamefont {P.}~\bibnamefont {B\"oni}},\ }\href {\doibase
  10.1103/PhysRevLett.102.186602} {\bibfield  {journal} {\bibinfo  {journal}
  {Phys. Rev. Lett.}\ }\textbf {\bibinfo {volume} {102}},\ \bibinfo {pages}
  {186602} (\bibinfo {year} {2009})}\BibitemShut {NoStop}%
\bibitem [{\citenamefont {Rashba}(1960)}]{Rashba}%
  \BibitemOpen
  \bibfield  {author} {\bibinfo {author} {\bibfnamefont {E.}~\bibnamefont
  {Rashba}},\ }\href@noop {} {\bibfield  {journal} {\bibinfo  {journal} {Sov.
  Phys.}\ }\textbf {\bibinfo {volume} {2}},\ \bibinfo {pages} {1109} (\bibinfo
  {year} {1960})}\BibitemShut {NoStop}%
\bibitem [{\citenamefont {Dresselhaus}(1955)}]{Dresselhaus}%
  \BibitemOpen
  \bibfield  {author} {\bibinfo {author} {\bibfnamefont {G.}~\bibnamefont
  {Dresselhaus}},\ }\href {\doibase 10.1103/PhysRev.100.580} {\bibfield
  {journal} {\bibinfo  {journal} {Phys. Rev.}\ }\textbf {\bibinfo {volume}
  {100}},\ \bibinfo {pages} {580} (\bibinfo {year} {1955})}\BibitemShut
  {NoStop}%
\bibitem [{\citenamefont {Rushforth}\ \emph {et~al.}(2007)\citenamefont
  {Rushforth}, \citenamefont {V\'yborn\'y}, \citenamefont {King}, \citenamefont
  {Edmonds}, \citenamefont {Campion}, \citenamefont {Foxon}, \citenamefont
  {Wunderlich}, \citenamefont {Irvine}, \citenamefont
  {Va\ifmmode~\check{s}\else \v{s}\fi{}ek}, \citenamefont {Nov\'ak},
  \citenamefont {Olejn\'{\i}k}, \citenamefont {Sinova}, \citenamefont
  {Jungwirth},\ and\ \citenamefont {Gallagher}}]{GaAsAMR}%
  \BibitemOpen
  \bibfield  {author} {\bibinfo {author} {\bibfnamefont {A.~W.}\ \bibnamefont
  {Rushforth}}, \bibinfo {author} {\bibfnamefont {K.}~\bibnamefont
  {V\'yborn\'y}}, \bibinfo {author} {\bibfnamefont {C.~S.}\ \bibnamefont
  {King}}, \bibinfo {author} {\bibfnamefont {K.~W.}\ \bibnamefont {Edmonds}},
  \bibinfo {author} {\bibfnamefont {R.~P.}\ \bibnamefont {Campion}}, \bibinfo
  {author} {\bibfnamefont {C.~T.}\ \bibnamefont {Foxon}}, \bibinfo {author}
  {\bibfnamefont {J.}~\bibnamefont {Wunderlich}}, \bibinfo {author}
  {\bibfnamefont {A.~C.}\ \bibnamefont {Irvine}}, \bibinfo {author}
  {\bibfnamefont {P.}~\bibnamefont {Va\ifmmode~\check{s}\else \v{s}\fi{}ek}},
  \bibinfo {author} {\bibfnamefont {V.}~\bibnamefont {Nov\'ak}}, \bibinfo
  {author} {\bibfnamefont {K.}~\bibnamefont {Olejn\'{\i}k}}, \bibinfo {author}
  {\bibfnamefont {J.}~\bibnamefont {Sinova}}, \bibinfo {author} {\bibfnamefont
  {T.}~\bibnamefont {Jungwirth}}, \ and\ \bibinfo {author} {\bibfnamefont
  {B.~L.}\ \bibnamefont {Gallagher}},\ }\href {\doibase
  10.1103/PhysRevLett.99.147207} {\bibfield  {journal} {\bibinfo  {journal}
  {Phys. Rev. Lett.}\ }\textbf {\bibinfo {volume} {99}},\ \bibinfo {pages}
  {147207} (\bibinfo {year} {2007})}\BibitemShut {NoStop}%
\bibitem [{\citenamefont {Dugaev}\ \emph {et~al.}(2005)\citenamefont {Dugaev},
  \citenamefont {Bruno}, \citenamefont {Taillefumier}, \citenamefont {Canals},\
  and\ \citenamefont {Lacroix}}]{AHE1}%
  \BibitemOpen
  \bibfield  {author} {\bibinfo {author} {\bibfnamefont {V.~K.}\ \bibnamefont
  {Dugaev}}, \bibinfo {author} {\bibfnamefont {P.}~\bibnamefont {Bruno}},
  \bibinfo {author} {\bibfnamefont {M.}~\bibnamefont {Taillefumier}}, \bibinfo
  {author} {\bibfnamefont {B.}~\bibnamefont {Canals}}, \ and\ \bibinfo {author}
  {\bibfnamefont {C.}~\bibnamefont {Lacroix}},\ }\href {\doibase
  10.1103/PhysRevB.71.224423} {\bibfield  {journal} {\bibinfo  {journal} {Phys.
  Rev. B}\ }\textbf {\bibinfo {volume} {71}},\ \bibinfo {pages} {224423}
  (\bibinfo {year} {2005})}\BibitemShut {NoStop}%
\bibitem [{\citenamefont {Sinitsyn}\ \emph {et~al.}(2005)\citenamefont
  {Sinitsyn}, \citenamefont {Niu}, \citenamefont {Sinova},\ and\ \citenamefont
  {Nomura}}]{AHE2}%
  \BibitemOpen
  \bibfield  {author} {\bibinfo {author} {\bibfnamefont {N.~A.}\ \bibnamefont
  {Sinitsyn}}, \bibinfo {author} {\bibfnamefont {Q.}~\bibnamefont {Niu}},
  \bibinfo {author} {\bibfnamefont {J.}~\bibnamefont {Sinova}}, \ and\ \bibinfo
  {author} {\bibfnamefont {K.}~\bibnamefont {Nomura}},\ }\href {\doibase
  10.1103/PhysRevB.72.045346} {\bibfield  {journal} {\bibinfo  {journal} {Phys.
  Rev. B}\ }\textbf {\bibinfo {volume} {72}},\ \bibinfo {pages} {045346}
  (\bibinfo {year} {2005})}\BibitemShut {NoStop}%
\bibitem [{\citenamefont {Mihai~Miron}\ \emph {et~al.}(2010)\citenamefont
  {Mihai~Miron}, \citenamefont {Gaudin}, \citenamefont {Auffret}, \citenamefont
  {Rodmacq}, \citenamefont {Schuhl}, \citenamefont {Pizzini}, \citenamefont
  {Vogel},\ and\ \citenamefont {Gambardella}}]{SOT1}%
  \BibitemOpen
  \bibfield  {author} {\bibinfo {author} {\bibfnamefont {I.}~\bibnamefont
  {Mihai~Miron}}, \bibinfo {author} {\bibfnamefont {G.}~\bibnamefont {Gaudin}},
  \bibinfo {author} {\bibfnamefont {S.}~\bibnamefont {Auffret}}, \bibinfo
  {author} {\bibfnamefont {B.}~\bibnamefont {Rodmacq}}, \bibinfo {author}
  {\bibfnamefont {A.}~\bibnamefont {Schuhl}}, \bibinfo {author} {\bibfnamefont
  {S.}~\bibnamefont {Pizzini}}, \bibinfo {author} {\bibfnamefont
  {J.}~\bibnamefont {Vogel}}, \ and\ \bibinfo {author} {\bibfnamefont
  {P.}~\bibnamefont {Gambardella}},\ }\href {\doibase 10.1038/nmat2613}
  {\bibfield  {journal} {\bibinfo  {journal} {Nat. Mater.}\ }\textbf {\bibinfo
  {volume} {9}},\ \bibinfo {pages} {230} (\bibinfo {year} {2010})}\BibitemShut
  {NoStop}%
\bibitem [{\citenamefont {Han}\ \emph {et~al.}(2016)\citenamefont {Han},
  \citenamefont {Zhang}, \citenamefont {Zhu}, \citenamefont {Du}, \citenamefont
  {Ge}, \citenamefont {Ling}, \citenamefont {Pi}, \citenamefont {Zhang},\ and\
  \citenamefont {Zhang}}]{CrGe1}%
  \BibitemOpen
  \bibfield  {author} {\bibinfo {author} {\bibfnamefont {H.}~\bibnamefont
  {Han}}, \bibinfo {author} {\bibfnamefont {L.}~\bibnamefont {Zhang}}, \bibinfo
  {author} {\bibfnamefont {X.}~\bibnamefont {Zhu}}, \bibinfo {author}
  {\bibfnamefont {H.}~\bibnamefont {Du}}, \bibinfo {author} {\bibfnamefont
  {M.}~\bibnamefont {Ge}}, \bibinfo {author} {\bibfnamefont {L.}~\bibnamefont
  {Ling}}, \bibinfo {author} {\bibfnamefont {L.}~\bibnamefont {Pi}}, \bibinfo
  {author} {\bibfnamefont {C.}~\bibnamefont {Zhang}}, \ and\ \bibinfo {author}
  {\bibfnamefont {Y.}~\bibnamefont {Zhang}},\ }\href
  {http://dx.doi.org/10.1038/srep39338} {\bibfield  {journal} {\bibinfo
  {journal} {Sci. Rep.}\ }\textbf {\bibinfo {volume} {6}} (\bibinfo {year}
  {2016})}\BibitemShut {NoStop}%
\bibitem [{\citenamefont {Ghimire}\ \emph {et~al.}(2012)\citenamefont
  {Ghimire}, \citenamefont {McGuire}, \citenamefont {Parker}, \citenamefont
  {Sales}, \citenamefont {Yan}, \citenamefont {Keppens}, \citenamefont
  {Koehler}, \citenamefont {Latture},\ and\ \citenamefont {Mandrus}}]{CrGe2}%
  \BibitemOpen
  \bibfield  {author} {\bibinfo {author} {\bibfnamefont {N.~J.}\ \bibnamefont
  {Ghimire}}, \bibinfo {author} {\bibfnamefont {M.~A.}\ \bibnamefont
  {McGuire}}, \bibinfo {author} {\bibfnamefont {D.~S.}\ \bibnamefont {Parker}},
  \bibinfo {author} {\bibfnamefont {B.~C.}\ \bibnamefont {Sales}}, \bibinfo
  {author} {\bibfnamefont {J.-Q.}\ \bibnamefont {Yan}}, \bibinfo {author}
  {\bibfnamefont {V.}~\bibnamefont {Keppens}}, \bibinfo {author} {\bibfnamefont
  {M.}~\bibnamefont {Koehler}}, \bibinfo {author} {\bibfnamefont {R.~M.}\
  \bibnamefont {Latture}}, \ and\ \bibinfo {author} {\bibfnamefont
  {D.}~\bibnamefont {Mandrus}},\ }\href {\doibase 10.1103/PhysRevB.85.224405}
  {\bibfield  {journal} {\bibinfo  {journal} {Phys. Rev. B}\ }\textbf {\bibinfo
  {volume} {85}},\ \bibinfo {pages} {224405} (\bibinfo {year}
  {2012})}\BibitemShut {NoStop}%
\bibitem [{\citenamefont {Zagryazhskii}\ \emph {et~al.}(1968)\citenamefont
  {Zagryazhskii}, \citenamefont {Gel'd},\ and\ \citenamefont
  {Shtol'ts}}]{CrGe3}%
  \BibitemOpen
  \bibfield  {author} {\bibinfo {author} {\bibfnamefont {V.~L.}\ \bibnamefont
  {Zagryazhskii}}, \bibinfo {author} {\bibfnamefont {P.~V.}\ \bibnamefont
  {Gel'd}}, \ and\ \bibinfo {author} {\bibfnamefont {A.~K.}\ \bibnamefont
  {Shtol'ts}},\ }\href {\doibase 10.1007/BF00816595} {\bibfield  {journal}
  {\bibinfo  {journal} {Sov. Phys. J.}\ }\textbf {\bibinfo {volume} {11}},\
  \bibinfo {pages} {23} (\bibinfo {year} {1968})}\BibitemShut {NoStop}%
\bibitem [{\citenamefont {Kolenda}\ \emph {et~al.}(1980)\citenamefont
  {Kolenda}, \citenamefont {Stoch},\ and\ \citenamefont {Szytu{\l}a}}]{CrGe4}%
  \BibitemOpen
  \bibfield  {author} {\bibinfo {author} {\bibfnamefont {M.}~\bibnamefont
  {Kolenda}}, \bibinfo {author} {\bibfnamefont {J.}~\bibnamefont {Stoch}}, \
  and\ \bibinfo {author} {\bibfnamefont {A.}~\bibnamefont {Szytu{\l}a}},\
  }\href@noop {} {\bibfield  {journal} {\bibinfo  {journal} {J. Magn. Magn.
  Mater.}\ }\textbf {\bibinfo {volume} {20}},\ \bibinfo {pages} {99} (\bibinfo
  {year} {1980})}\BibitemShut {NoStop}%
\bibitem [{\citenamefont {Nagaosa}\ \emph {et~al.}(2010)\citenamefont
  {Nagaosa}, \citenamefont {Sinova}, \citenamefont {Onoda}, \citenamefont
  {MacDonald},\ and\ \citenamefont {Ong}}]{Nagaosa2010}%
  \BibitemOpen
  \bibfield  {author} {\bibinfo {author} {\bibfnamefont {N.}~\bibnamefont
  {Nagaosa}}, \bibinfo {author} {\bibfnamefont {J.}~\bibnamefont {Sinova}},
  \bibinfo {author} {\bibfnamefont {S.}~\bibnamefont {Onoda}}, \bibinfo
  {author} {\bibfnamefont {A.~H.}\ \bibnamefont {MacDonald}}, \ and\ \bibinfo
  {author} {\bibfnamefont {N.~P.}\ \bibnamefont {Ong}},\ }\href {\doibase
  10.1103/RevModPhys.82.1539} {\bibfield  {journal} {\bibinfo  {journal} {Rev.
  Mod. Phys.}\ }\textbf {\bibinfo {volume} {82}},\ \bibinfo {pages} {1539}
  (\bibinfo {year} {2010})}\BibitemShut {NoStop}%
\bibitem [{\citenamefont {Tian}\ \emph {et~al.}(2009)\citenamefont {Tian},
  \citenamefont {Ye},\ and\ \citenamefont {Jin}}]{properscalingAHE}%
  \BibitemOpen
  \bibfield  {author} {\bibinfo {author} {\bibfnamefont {Y.}~\bibnamefont
  {Tian}}, \bibinfo {author} {\bibfnamefont {L.}~\bibnamefont {Ye}}, \ and\
  \bibinfo {author} {\bibfnamefont {X.}~\bibnamefont {Jin}},\ }\href {\doibase
  10.1103/PhysRevLett.103.087206} {\bibfield  {journal} {\bibinfo  {journal}
  {Phys. Rev. Lett.}\ }\textbf {\bibinfo {volume} {103}},\ \bibinfo {pages}
  {087206} (\bibinfo {year} {2009})}\BibitemShut {NoStop}%
\bibitem [{\citenamefont {Masuda}\ \emph {et~al.}(1977)\citenamefont {Masuda},
  \citenamefont {Hioki},\ and\ \citenamefont {Oota}}]{NMR}%
  \BibitemOpen
  \bibfield  {author} {\bibinfo {author} {\bibfnamefont {Y.}~\bibnamefont
  {Masuda}}, \bibinfo {author} {\bibfnamefont {T.}~\bibnamefont {Hioki}}, \
  and\ \bibinfo {author} {\bibfnamefont {A.}~\bibnamefont {Oota}},\ }\href
  {\doibase http://dx.doi.org/10.1016/0378-4363(77)90197-8} {\bibfield
  {journal} {\bibinfo  {journal} {Physica B+C}\ }\textbf {\bibinfo {volume}
  {91}},\ \bibinfo {pages} {291 } (\bibinfo {year} {1977})}\BibitemShut
  {NoStop}%
\bibitem [{\citenamefont {McGuire}\ and\ \citenamefont {Potter}(1975)}]{AMR3d}%
  \BibitemOpen
  \bibfield  {author} {\bibinfo {author} {\bibfnamefont {T.}~\bibnamefont
  {McGuire}}\ and\ \bibinfo {author} {\bibfnamefont {R.}~\bibnamefont
  {Potter}},\ }\href {\doibase 10.1109/TMAG.1975.1058782} {\bibfield  {journal}
  {\bibinfo  {journal} {IEEE T. Magn.}\ }\textbf {\bibinfo {volume} {11}},\
  \bibinfo {pages} {1018} (\bibinfo {year} {1975})}\BibitemShut {NoStop}%
\bibitem [{\citenamefont {van Gorkom}\ \emph {et~al.}(2001)\citenamefont {van
  Gorkom}, \citenamefont {Caro}, \citenamefont {Klapwijk},\ and\ \citenamefont
  {Radelaar}}]{AMRFe}%
  \BibitemOpen
  \bibfield  {author} {\bibinfo {author} {\bibfnamefont {R.~P.}\ \bibnamefont
  {van Gorkom}}, \bibinfo {author} {\bibfnamefont {J.}~\bibnamefont {Caro}},
  \bibinfo {author} {\bibfnamefont {T.~M.}\ \bibnamefont {Klapwijk}}, \ and\
  \bibinfo {author} {\bibfnamefont {S.}~\bibnamefont {Radelaar}},\ }\href
  {\doibase 10.1103/PhysRevB.63.134432} {\bibfield  {journal} {\bibinfo
  {journal} {Phys. Rev. B}\ }\textbf {\bibinfo {volume} {63}},\ \bibinfo
  {pages} {134432} (\bibinfo {year} {2001})}\BibitemShut {NoStop}%
\bibitem [{\citenamefont {Smit}(1951)}]{AMRSmit}%
  \BibitemOpen
  \bibfield  {author} {\bibinfo {author} {\bibfnamefont {J.}~\bibnamefont
  {Smit}},\ }\href {\doibase http://dx.doi.org/10.1016/0031-8914(51)90117-6}
  {\bibfield  {journal} {\bibinfo  {journal} {Physica}\ }\textbf {\bibinfo
  {volume} {17}},\ \bibinfo {pages} {612 } (\bibinfo {year}
  {1951})}\BibitemShut {NoStop}%
\end{thebibliography}%

\end{document}